\documentclass{emulateapj}
\usepackage{float}
\usepackage{appendix}
\usepackage{mathtools}
\usepackage{epsfig}
\usepackage{array}
\usepackage{graphicx}
\usepackage{graphics}
\usepackage{subfigure}
\usepackage[latin1]{inputenc}
\usepackage{latexsym}

\shorttitle{Cold Dust Around Fast-expanding SNe Ia}
\shortauthors{Xiaofeng Wang et al.}

\def\gsim{\;\lower4pt\hbox{${\buildrel\displaystyle >\over\sim}$}\;}
\def\lsim{\;\lower4pt\hbox{${\buildrel\displaystyle <\over\sim}$}\;}
\def\grls{\;\lower4pt\hbox{${\buildrel\displaystyle >\over <}$}\;}

\begin{document}

\title{The Cold and Dusty Circumstellar Matter around \\ Fast-expanding Type Ia Supernovae}
\author{
Xiaofeng~Wang\altaffilmark{1}, Jia Chen\altaffilmark{1},
Lifan Wang\altaffilmark{2,3}, Maokai~Hu\altaffilmark{3},
Gaobo Xi\altaffilmark{1}, Yi Yang\altaffilmark{4,2},
Xulin Zhao\altaffilmark{5,1}, Wenxiong Li\altaffilmark{1},
}

\altaffiltext{1}{Physics Department and Tsinghua Center for Astrophysics, Tsinghua University, Beijing 100084, China; wang\_xf@mail.tsinghua.edu.cn}
\altaffiltext{2}{Mitchell Institute for Fundamental Physics and Astronomy, Texas A\&M University, College Station, TX 77843, USA}
\altaffiltext{3}{Purple Mountain Observatory, Nanjing, 201008, Jiangsu, China}
\altaffiltext{4}{Department of Particle Physics and Astrophysics, Weizmann Institute of Science, Rehovot 76100, Israel}
\altaffiltext{5}{Department of Physics, Tianjin University of Technology, Tianjin 300384, China}

\begin{abstract}

Type Ia supernovae (SNe~Ia) play key roles in revealing the accelerating expansion of the universe, but our knowledge about their progenitors is still very limited. Here we report the discovery of a rigid dichotomy in circumstellar (CS) environments around two subclasses of type Ia supernovae (SNe Ia) as defined by their distinct photospheric velocities. For the SNe Ia with high photospheric velocities (HV), we found a significant excess flux in blue light during 60-100 days past maximum, while this phenomenon is absent for SNe with normal photospheric velocity (Normal). This blue excess can be attributed to light echoes by circumstellar dust located at a distance of about 1-3$\times$10$^{17}$ cm from the HV subclass. Moreover, we also found that the HV SNe Ia show systematically evolving Na~{\sc I} absorption line by performing a systematic search of variable Na~{\sc I} absorption lines in spectra of all SNe Ia, whereas this evolution is rarely seen in Normal ones. The evolving Na~{\sc I} absorption can be modeled in terms of photoionization model, with the location of the gas clouds at a distance of about 2$\times$10$^{17}$ cm, in striking agreement with the location of CS dust inferred from B-band light curve excess. These observations show clearly that the progenitors of HV and Normal subclasses are systematically different, suggesting that they are likely from single and double degenerate progenitor systems, respectively.

\end{abstract}


\section{Introduction}
\label{sect:intro}
It is conventionally accepted that type Ia supernovae (SNe Ia) result from thermonuclear explosion of a carbon-oxygen (CO) white dwarf (Nomoto et al. 1997, Hillebrandt et al. 2000, Maoz 2014). Two popular scenarios are: merger-induced explosion of two white dwarfs (the so-called double degenerate (DD) scenario) and accretion-induced explosion of a massive WD with a non-degenerate companion (the so-called single degenerate (SD) scenario). Many progenitor classes have been proposed (Gilfanov et al. 2010, Wang \& Han 2012, Maoz 2014), but observational evidences have not yet reached definitive conclusions on particular progenitor systems. The SD scenario is favored by the possible detections of circumstellar materials (CSM) around some SNe Ia through detections of strong ejecta-CSM interaction (Hamuy et al. 2003, Wang et al. 2004, Aldering et al. 2006, Taddia et~al. 2012, Silverman et al. 2013, Bochenek et al. 2018) or evolving narrow absorption lines possibly due to CSM (Patat et al. 2007, Blondin et al. 2009, Sternberg et al. 2011, Dilday et al. 2012), while there are also observational findings suggesting no companion signatures for some SNe~Ia (Li et al. 2011a, Gonzalez et al. 2012, Schaefer et al. 2012, Olling et al. 2015). This may suggest that SNe Ia have multiple progenitor systems, as favored by the discovery that SNe Ia with different ejecta velocities originate from distinct birthplace environments (Wang et al. 2013).

Observationally, spectroscopically normal SNe~Ia consists of $\sim$70\% of all SNe~Ia (Branch et al. 1993, Li et al. 2011b) and can be categorized into high-velocity (HV) and normal-velocity (NV) subclasses based on ejecta velocities inferred from the blueshifted Si~II 6355 \AA\ line, i.e., 12,000 km s$^{-1}$ (Wang et al. 2009). Compared to NV ones, HV SNe~Ia have systematically redder $B - V$ colors at maximum light and prefer abnormally low total to selective absorption R$_{V}$ ratios (Wang et al. 2009). It is thus important to examine whether this difference arises from the intrinsic difference of the SN ejecta or from systematic difference in circumstellar (CS) and/or interstellar environments of the two groups. These issues may shed light on the elusive progenitor systems of SNe~Ia.

The presence of CS dust can be tested by examining the behaviors of narrow interstellar absorption features. The interstellar sodium Na~{\sc I} doublet (D$_1$ 5896~\AA, D$_2$ 5890~\AA) is a good tracer of gas, dust and metals, and its strength is found to show positive correlation with line-of-sight dust reddening (Munari \& Zwitter 1997, Turatto et al. 2003, Blondin et al. 2009, Folatelli et al. 2010, Pozanski et al. 2012, Phillips et al. 2013). Its evolution (or variation) and velocity structure can provide significant constraints on the presence and distance of CS dust around SNe~Ia (Patat et al. 2007, Chugai 2008, Simon et al. 2009, Sternberg et al. 2011), while it is not clear why evolution or velocity structure of interstellar Na~I lines exist in some SNe Ia but not in the others. On the other hand, the surrounding CS dust may also affect the light curves of SNe Ia as a result of light scattering by the nearby dust. This can be examined by inspecting light curves in the early nebular phase.

In this paper, we conducted a systematic search for variable Na~{\sc I} absorption in low-resolution spectra of SNe Ia as well as an overall analysis of the behavior of their late-time light curves, with an attempt to constrain the dust environments around SNe Ia and hence the properties of their progenitors. This paper is organized as follows: in Section 2, we describe the dataset of SNe Ia used in our analysis. The results from interstellar Na absorptions and late-time light curves are presented in Section 3. Discussions and conclusions are given in Section 4.

\section{Dataset}
The spectral sample used to measure the Na absorption features in SNe Ia are primarily from the Center for Astrophysics (CfA) Supernova Program (Riess et al. 1996), the Carnegie Supernova Project (CSP; Hamuy et al. 1996). The former sample contains 2603 spectra of 462 nearby SNe Ia (Blondin et al. 2012), and most (94\%) of which were obtained with the FAST spectrograph on the 1.5~m telescope at the Fred Lawrence Whipple Observatory (FLWO). The latter dataset contains 604 spectra of 93 SNe~Ia (Folatelli et al. 2013), which were mainly obtained with the 2.5~m du Pont Telescope at Las Campanas Observatory. All of the spectra were reduced in a consistent way and have a typical FWHM (full-width at half maximum) resolution of 6-7 {\AA}, providing the largest homogeneous spectroscopic dataset of SNe Ia. We also used the spectra from the Berkeley Supernova Program (BSP, Silverman et al. 2012), consisting of 1298 spectra for 582 SNe Ia, to get further classifications of our SN Ia sample and measure the Na absorptions whenever necessary. The spectral phases were obtained with respect to the $B$-band maximum light, based on the published light curves of CfA (Riess et al. 1999, Jha et al. 2006, Hicken et al. 2009, Hicken et al. 2012), Lick Observatory Supernova Survey (LOSS, Ganeshalingam et al. 2010), and Carnegie Supernova Project (Contreras et al. 2010, Stritzinger et al. 2011, Krisciunas et al. 2017). The light-curve parameters, including the peak magnitudes, the $B_{max} - V_{max}$ colors at the maximum light, and the post-maximum decline rates $\Delta m_{15}$(B) (Phillips 1993) are estimated by applying polynomial fits and/or the SALT2 fit (Guy et~al. 2007) to the observed data. A weighted average is adopted when the estimations from the above two methods are both reasonable. The measurement results and relevant photometric parameters for each SN sample are listed in Table~1.

To eliminate the effects from noise spikes on the continuum determination of Na~I absorption doublet, we smoothed the observed spectra in the range of 5850$\sim$5950~\AA\ (after corrections for the redshift of host galaxies) using a gaussian function with an FWHM ($\sim$2.35 sigma) of 7 Angstrom. This amount is chosen because it is comparable to the typical width of the Na I absorption doublet in the spectra used in our analysis and this match can also optimize the resolution of signal from noise according to the matched filter theorem. The smoothed spectra are then used to find the local flux maximum and hence define the pseudo-continuum. The definition of pseudo-continuum and hence the calculations of the EW are automatic for our SN sample, but we also double-checked the integration limits by eyes to make sure that they are reasonably determined. For a few SNe whose spectra suffering from improper wavelength calibration and/or motion of the SNe relative to the galaxy center (e.g., SN 1998eg and SN 2007ai), we adjusted their integration limit manually. As the Na~{\sc I}D absorption of some SNe~Ia is found to show temporal evolution within one week from the maximum light (see also Figure 2), the EW of each SN is taken as a weighted mean of the results obtained at t$\sim$0-30 days after the maximum light when multi-epoch spectra are available.

As late-time photometric evolution can provide an important probe of dust environments around SNe~Ia, we thus also include sample with better photometric observations in the early nebular phase even though they may not have good spectral data allowing for accurate measurement of Na I~D absorption. The total sample thus contains 206 SNe Ia, including subtypes of 112 NV, 54 HV, 25 91T-like, and 15 91bg-like objects, which corresponds to a percentage of 54.4\%, 26.2\%, 12.1\%, and 7.3\%, respectively. These fractions of different subtypes are basically consistent with the ones obtained with a magnitude-limited sample of SNe Ia (i.e., Li et~al. 2011b).

Table 1 lists the relevant parameters of our SN Ia sample, and the meaning of each column is as follows: Column (1), SN name; Column (2), the SN~Ia subtype based on the Si~II velocity at around the maximum light (Wang et al. 2009); Column (3), ejecta velocity measured from Si~II 6355 absorption in the near-maximum-light spectra, estimated using multiple-gaussian technique described in Zhao et~al. (2015); Column (4),$\Delta m_{15}$(B), the B-band magnitude decline rate in 15 days from the maximum ; Column (5), $B_{max} - V_{max}$ color at the maximum light, corrected for the Galactic reddening; Column (6), the weighted mean value of the equivalent width (EW) of Na I absorption over the period 0 $\lesssim$ t $\lesssim$ 30 days whenever possible; Column (7), the $B$-band magnitude decline in the first 60 days after the B-band maximum; Column (8), the $V$-band magnitude decline in the first 60 days after the B-band maximum; Column (9), References.

\section{Results}
\subsection{Narrow Na {\sc I} Absorption Features}
Figure~1 shows the $B - V$ colors at the maximum light versus the EWs of the Na {\sc I}D absorption due to the host galaxies. Foreground Galactic reddening corrections have been removed from the observed colors. The $B_{max} - V_{max}$ color has been usually used as an indicator of reddening (Phillips et al. 1999, Wang et al. 2009, Folatelli et al. 2010, and Burns et al. 2014), though the subclasses of 91T-like and 91bg-like SNe~Ia tend to have peculiar colors and they are thus not shown in Fig.1.

\begin{figure}
\vspace{-0.5cm}
\hspace{-0.5cm}
\includegraphics[width=100mm, height=70mm]{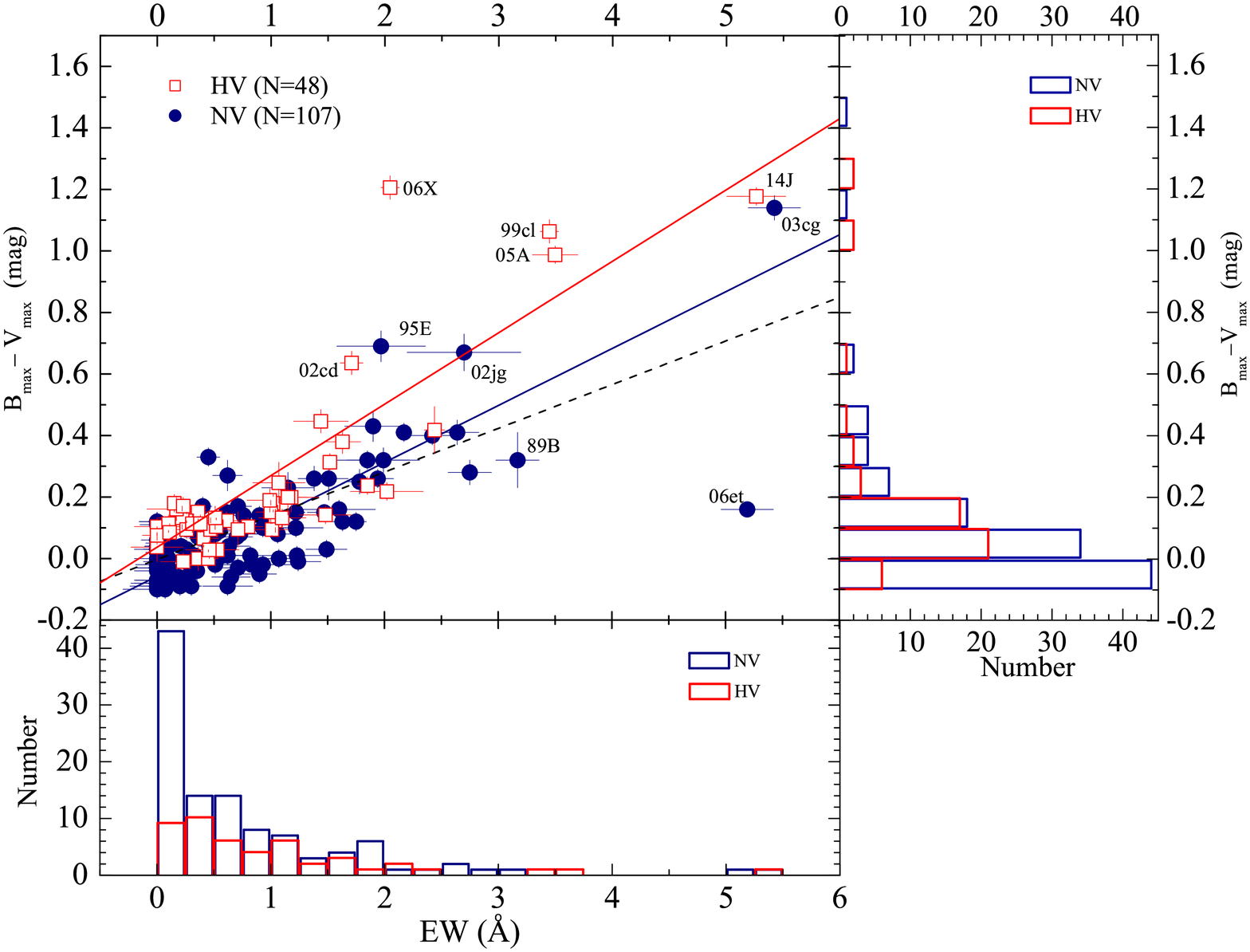} \\
\vspace{-0.8cm}
\caption{Relation between absorption strength (expressed as equivalent width) of interstellar sodium lines in spectra of SNe Ia and the corresponding values of their $B_{max}-V_{max}$ colors. The "HV" subgroup (red squares) represents the SNe Ia with larger Si~II velocity at $B$-band maximum light, i.e., v$^{0}$$_{Si~II}$$\geq$ 12,000 km s$^{-1}$, while the "NV" subgroup (blue circles) denotes those with v$^{0}_{Si~II}$ $<$ 12,000 km s$^{-1}$. The red and blue lines represent the best-fit linear relationship to the HV and NV SNe Ia, respectively, while the dashed, black line shows the relationship inferred from the Na absorption and reddening due to the Milky Way. SN 2006X was not included in the fit due to the saturation of the Na~I lines (Patat et al. 2007), and SN 2006et appears to be an outlier and was not included in the fit, either. The plot is then projected onto the two side panels where a histogram is displayed for each SN Ia subclass in each of the dimensions.}
\end{figure}

As can be readily seen, the correlation between $B_{max} - V_{max}$ color and strength of Na absorption is apparent for the HV and NV subclasses, with stronger Na {\sc I} D absorptions corresponding to redder colors (and hence larger reddening). However, these two subclasses of SNe Ia show noticeable differences in the distribution of their $B_{max} - V_{max}$ colors and the Na {\sc I}D EWs. A two-dimensional Kolmogoroff-Smirnoff (KS) test gives a probability of 0.01\% that they have statistically identical Na {\sc I}D EW and color distribution. Examining the EW of Na {\sc I} absorption and the $B_{max} - V_{max}$ color separately yields a respective probability of 1.4\% and 0.0002\% that the two subclasses have the same parent distribution. The mean $B_{max}-V_{max}$ colors estimated for these two subclasses are 0.26 mag (HV) and 0.08 mag (NV), respectively, while the mean Na {\sc I}D EWs are 1.1 {\AA} (HV) and 0.7 {\AA} (NV), respectively. In general, the HV subclass have redder colors and stronger Na~{\sc I}D absorption lines, and the fraction found with EWs $\gtrsim$ 1.0 is as high as 44.8\% (versus 26.2\% for the NV counterparts). These results indicates that the HV SNe~Ia suffer systematically larger reddening on average than the NV ones; the intrinsic color of HV SNe~Ia may not necessarily be redder than their NV counterparts despite of their observed red color contrary to previous studies (i.e., Foley \& Kasen 2011).

Applying a linear fit to the $EW$ $-$ ($B_{max} - V_{max}$) relation for the HV and NV SNe Ia separately, we found that the former subclass have a larger slope of 0.239$\pm$0.016, while the slope for the latter is 0.177$\pm$0.011. This difference may be explained with incomplete recombination of the ionized Na (see Figure 2) and/or time evolving scattering of the SN photons by the surrounding dust (Wang 2005).

It is important to further constrain the location of the dust that gives rise to the excess extinction and the time varying Na {\sc I}D absorption in HV SNe~Ia. If the SN exploded near a dusty shell/slab, we may expect variable Na~{\sc I} absorptions in SN spectra due to photoionization effect (Chugai et al. 2008, Patat et al. 2011). The strength of the absorption may then be correlated with the rise and fall of the supernova luminosity. The trend that the Na~{\sc I}D absorptions become stronger after maximum light has already been reported for a few SNe~Ia such as SN 2006X (Patat et al. 2007), SN 2007le (Simon et al. 2009), SN 1999cl (Blondin et al. 2009), but the decrease in strength of Na~{\sc I} absorption expected shortly after explosion was never observed in SNe~Ia.

To investigate the EW variations, we consider all objects having at least three-epochs of spectroscopic data within $\sim$1 month from the maximum light. Applying a 3-$\sigma$ detection cut and a minimum EW variation of 0.5~{\AA} to the CfA and CSP data, we get a sample of 16 SNe~Ia showing prominent Na~{\sc I} D variations. Table 2 lists the relevant parameters of this sample showing variable Na absorptions. Of this 16 SNe~Ia, there are 11 HV and 2 NV SNe~Ia, with a fraction of 68.8\% and 12.5\%, respectively. This sample increases to 23 when applying a 2-$\sigma$ cut, which contains 13 HV SNe Ia (56.5\%) and 6 NV SNe Ia (26.0\%). The above analysis indicates that the varying of Na~{\sc I}D absorptions are statistically associated with the HV subclass. The probability of a chance coincidence is less than 0.1\%. Variable sodium features have already been detected in the high-resolution spectra of SN 2006X and SN 2007le, and low-resolution spectra of SN 1999cl. It is gratifying to note that the evolutions of their Na~{\sc I}D line are nicely recovered in these low resolution data.

Note that SN 2006dd was also proposed as an SN Ia showing significant changes in the Na I D absorptions (Stritzinger et al. 2010) and is thus listed in Table 2. Its EW of Na I D absorption seems to decrease from $\sim$3.2\AA at t$\sim$$-$12 and t$\sim$ +86 days to $\sim$1.3\AA at t$\sim$+137 days and then increase to $\sim$3.8\AA on +194 days. Owing to the lack of near-maximum-light spectra, we cannot put this object into HV or NV subclasses. We caution, however, that the center wavelength of the absorption features around 5890 Angstrom in the spectra of SN 2006dd changes from 5884 Angstrom in the +86.4 day spectrum to 5896 Angstrom in the t$\sim$+194.4 day spectrum. Such a large change may indicate improper wavelength calibration, or the identified "Na I" absorption might be actually caused by other unknown features given that this SN has relatively blue $B - V$ color.

\begin{figure}[htbp]
\center
\subfigure{%
\begin{minipage}[b]{1.\linewidth}
\centerline{\includegraphics[width=.99\linewidth]{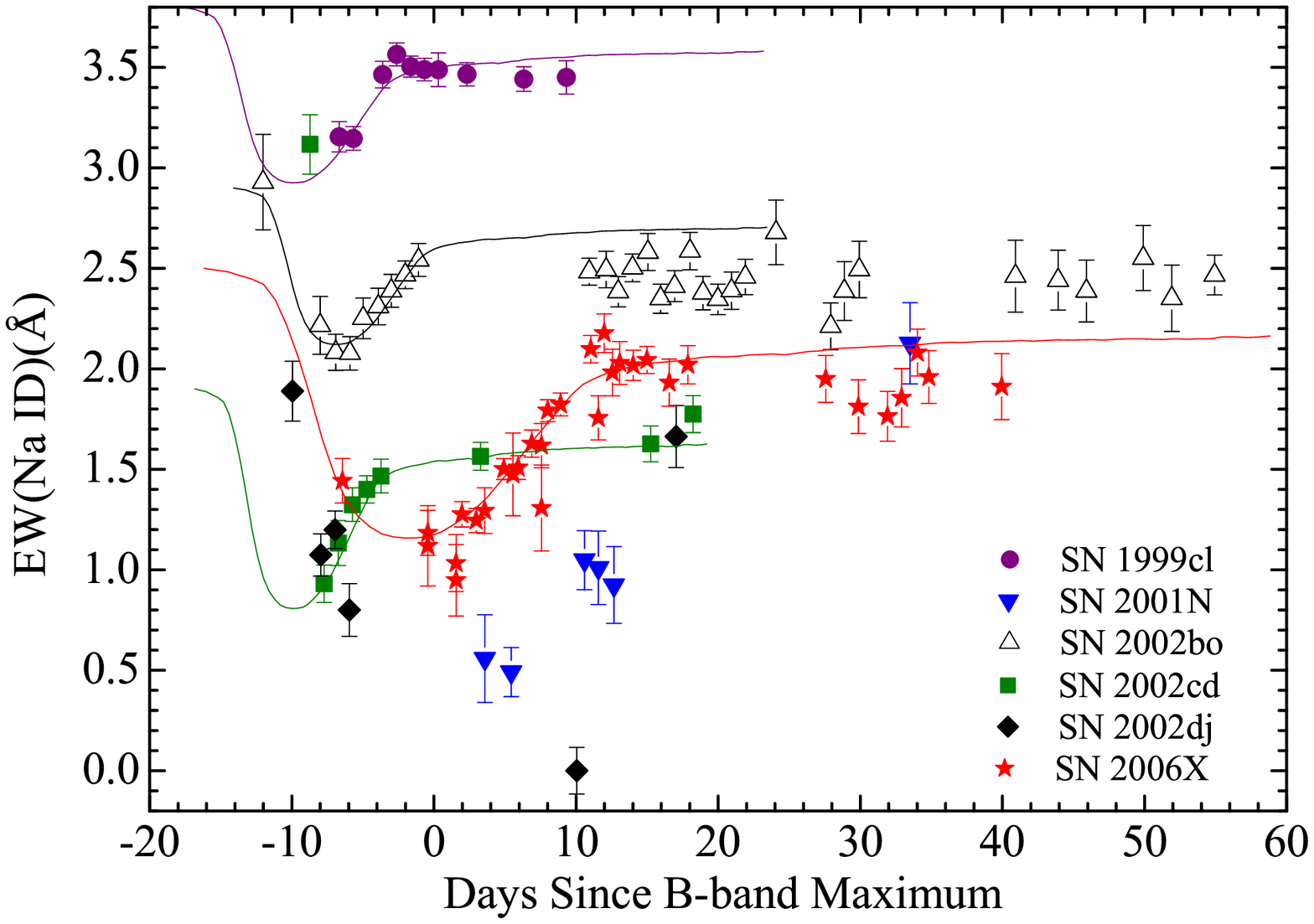}}
\end{minipage}}

\vspace{-1.0cm}
\subfigure{%
\begin{minipage}[b]{1.\linewidth}
\centerline{\includegraphics[width=.99\linewidth]{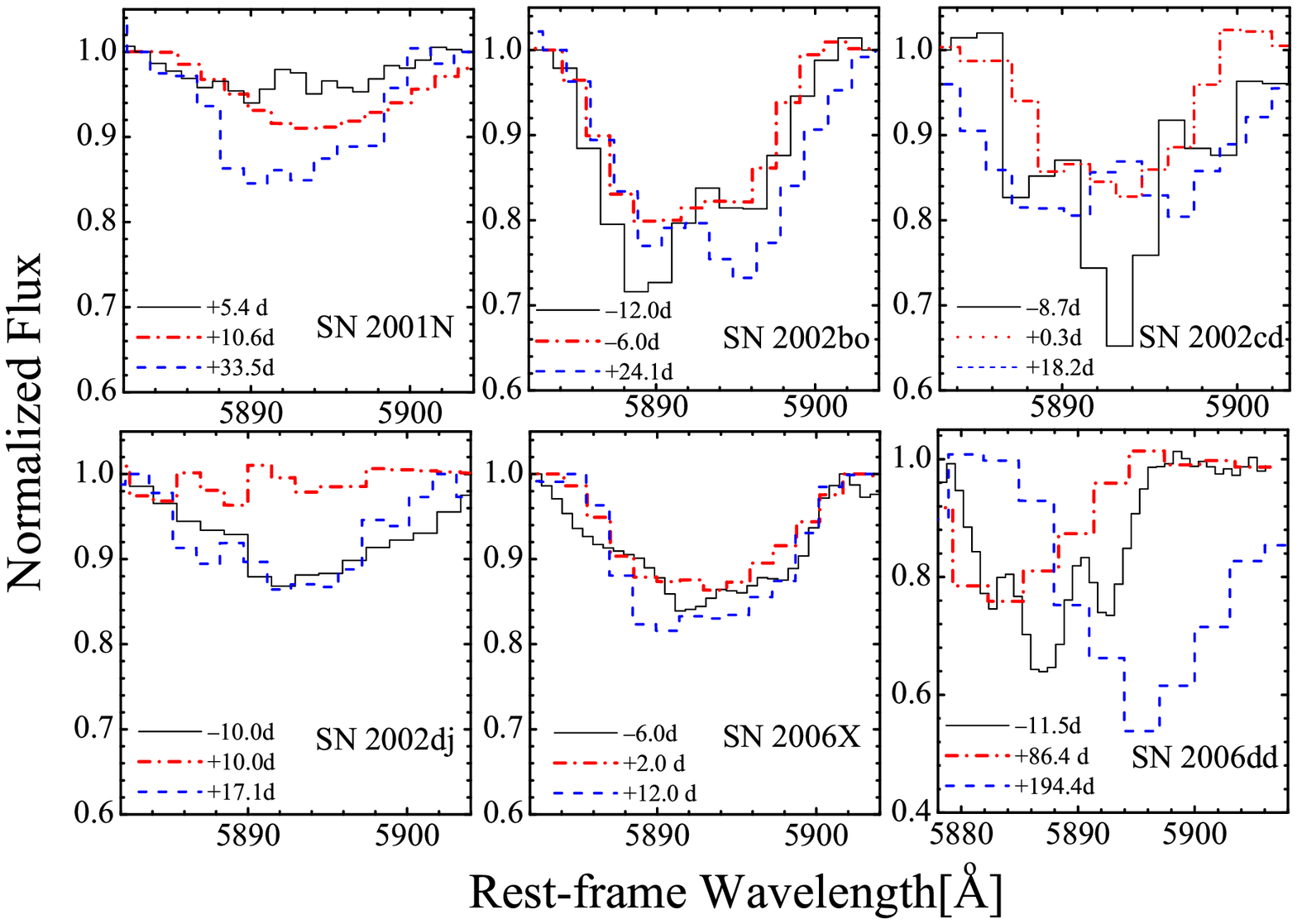}}
\caption{Temporal evolution of the equivalent width of Na~I doublet for type Ia supernovae (SNe Ia). Top panels: Variable Na absorptions detected in some well-observed objects such as SNe 1999cl, 2001N, 2002bo, 2002cd, 2002dj, and 2006X. Note that all of the 6 SNe Ia shown in the plot belong to the "HV" subclass. Overplotted are the curves derived with a model of Na~I photoionization and recombination in CS dust. Bottom: the continuum-normalized Na~I D profile shown at several epochs for some well-observed HV SNe Ia. The black and red lines represent the earlier and near-maximum-light phase spectra, respectively, while the blue ones show the profile in the recombination phase.}
\end{minipage}}
\end{figure}

The Na~{\sc I}D absorption of some representative SNe~Ia with significant time evolution is shown in Figure~2. The plot suggests that the Na~{\sc I}D line of these objects generally follow a qualitatively similar evolutionary trend. At early times the strength of the Na~{\sc I}D absorption was found to decrease with time, as is clearly seen in SN 2002bo, SN 2002dj, SN 2006X, and perhaps in SN 2002cd. Such an evolution can be attributed to photoionization of neutral Na by the UV photons of supernovae (Borkowski et al. 2009). After the declining phase, the Na absorption strength undergoes a rising stage lasting for about 10 days and remains at constant level after that. The overall evolution looks like a "square-root" sign, which can be well explained with a model of Na~{\sc I}D photoionization and recombination in CS dust (see solid curves in Figure 2) as discussed in \S 3.3.

\subsection{Late-time Light Curves}
The presence of CS dust can be also tested by examining the behaviors of late-time light curves of SNe Ia. This is because some of the SN photons will be scattered by surrounding dust and arrive at the observer with a time delay (Wang et al. 1996, Wang et al. 2005, Patat et al. 2006). The delayed photons can be seen as a light echo (LE) and may contribute to the observed light curves in the early nebular phase, especially in blue bands. To examine the difference of the late-time light curves for our sample of SNe Ia, we measured the magnitude decline within 60 days from the $B$- and $V$-band maximum light, as defined by $\Delta$m$_{60}$(B) and $\Delta$m$_{15}$(V), respectively. These two quantities are obtained by applying a linear fit to the observed data spanning the phase from t$\sim$+40 days to t$\sim$100 days, as listed in Table 1.

\begin{figure*}
\vspace{-1.8cm}
\hspace{0.5cm}
\figurenum{3}
\includegraphics[angle=0,width=170mm]{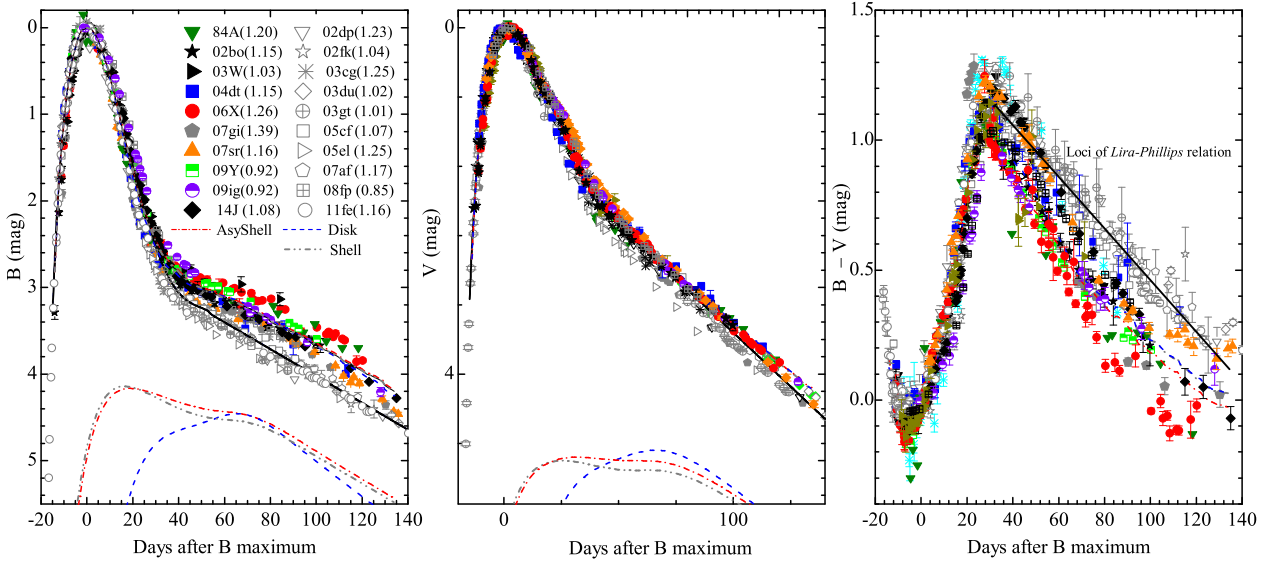}
\vspace{-2.0cm}
\caption{The $B$-(left panel), $V$-band light (middle panel) and $B - V$ color (right panel) curves of some well-observed SNe Ia (see Table 1 for the references of the data), with the curves being all normalized to the corresponding values at around the maximum light. The filled and semi-filled symbols the HV SNe Ia, and the open symbols denote the NV ones. SNe Ia of these two subclasses exhibit large differences in $B$ band and $B-V$ colors 40 days after the maximum brightness. The black solid lines indicate the mean light curves of NV subsample. The extra emission by scattering of SN light on CS materials of a spherical shell, asymmetric shell, and a disk structure, at a distance of $\sim$1-3$\times$ 10$^{17}$ cm, are shown with the gray dash-dot-dotted line, red dash-dotted, and blue dashed line, respectively. The resultant light curves by including contributions of CS scattering are indicated by the same symbols. The black solid line shows the best-fit to the $B - V$ color curves of NV SNe Ia using the updated $Lira-Phillips$ relation (Burns et~al. 2014), with the non-reddening loci being shifted redwards by 0.2 mag. During 40 days $<$ t $<$ 100 days, the color curves of most HV SNe Ia are apparently bluer than those of normal SNe Ia, following a slope much steeper than the $Lira-Phillips$ relation.} \label{fig-3}
\end{figure*}

Figure~3 shows the $BV$-band light curves and the $B-V$ color curves for some well-observed SNe~Ia. Although these SNe~Ia have similar light curves in early phases, they show remarkable scatter in the $B$-band from t$\sim$40 to 100 days after the peak. For instance, the $B$-band magnitude measured at t$\sim$60 days from the peak can differ by over 0.5 mag in the $B$ band (see Fig~3(a)). Such a late-time discrepancy is much smaller in the $V$ band (see Fig~3(b)), which leads to an apparently bluer $B - V$ color for the HV SNe~Ia relative to the NV objects, as shown in the right panel of Figure~3.  for

As more luminous SNe Ia tend to have brighter tails with slower decay rates, it is thus necessary to examine whether the excess emission seen in the tail light curves of HV SNe Ia is partially due to that they have intrinsically high luminosity. We plot the measured values of $\Delta$m$_{60}$ as a function of the corresponding $\Delta$m$_{15}$(B) for our sample in Figure 4, where one can see that the tail brightness does show a significant correlation with the decline rate in both $B$ and $V$ bands for the NV sample of SNe Ia. While this correlation shows large scatter in the $B$ band for the HV subsample, with the measured $\Delta$m$_{60}$(B) being systematically smaller than that of the NV ones at a given $\Delta$m$_{15}$(B). This comparison further confirms that the excess emission in the blue band is not intrinsic to the HV SNe Ia. Instead, we found that the $\Delta$m$_{60}$(B) shows a strong positive correlation with the Si~II 6355 velocity measured around the maximum light, as shown in Fig~4(c). This indicates that the origin of the larger expanding velocity might be closely related to the formation of dusty environments around SNe Ia.

\begin{figure*}
\figurenum{4}
\vspace{-1.0cm}
\hspace{-1.5cm}
\includegraphics[angle=0,width=200mm]{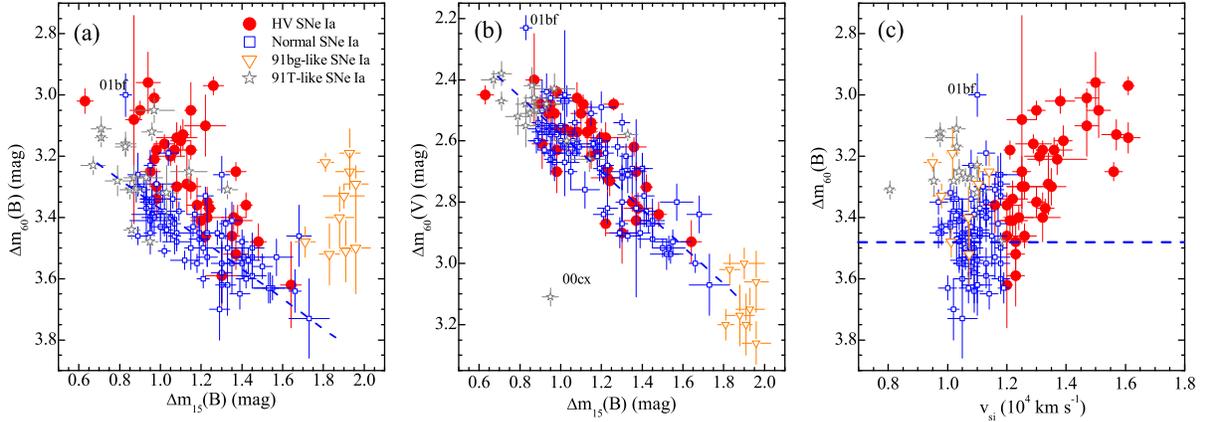}\\
\vspace{-7.0cm}
\caption{{\bf a,b}, Tail brightness of SNe Ia, measured as magnitude decline at t$\sim$+60 days from the peaks of $B$- and $V$-band light curves, versus the luminosity indicator of SNe Ia $\Delta m_{15}$(B) that is measured as the magnitude decline within the first 15 days after the $B$-band maximum (Phillips et al. 1993). {\bf c}, A plot of the $B$-band tail brightness as a function of Si~II velocity obtained around $B$-band maximum light. The subclasses of HV, NV, spectroscopically peculiar ones like SN 1991T and SN 1991bg SNe are shown by blue open squares, red dots, gray stars, and orange triangles, respectively. The blue dashed line in each panel represents the best linear fit for the NV subsample.}
\label{fig-4}
\end{figure*}

\subsection{Modeling of the Observed Results}
In this subsection, we will explore the presence of CS dust and constrain its distance from the SNe by comparing the modeling results with both the variations of Na I~D absorption observed in the spectra and the excess emission detected in the late-time light curves.

\subsubsection{Photonionization Model}
To study the effect of photonionizations on the observed Na~I absorptions in SN Ia spectra, we consider a simple model with a spherical CS dust shell. Similar models have been employed in previous studies (Borkowski et~al. 2009, Patat et~al. 2007) to study the variations in Na absorption lines, while only the re-combination process was emphasized. The Na ionization state is actually determined by the balance between ionizations and recombinations. In our calculations, both of these two stages are explored based on the change of the ionization and recombination rates with the ultraviolet flux emitted by the SNe Ia at different phases. Assuming that the pre-explosion number of Na~I in the CS shell is $N$(Na~I)(t$_{0}$) well before the SN explosion, and the change of this number over a time span $dt$ can be written as:

\begin{equation}
dN(Na~I)(t)/dt =\frac{dN(Na~I)_{ion}(t)+dN(Na~I)_{rec}(t)}{dt} \\
= -N(Na~I)(t_{0})\cdot R_{ion}(t)+[N(Na~I)-N(Na~I)(t)]\cdot R_{rec}(t).
\end{equation}

Where R$_{ion}$(t) and R$_{rec}$(t) represent the ionization rate of Na~I and recombination rate of Na~II, respectively.
Defining the fraction of Na~I as Frac(Na~I(t))=$\frac{N(Na~I)(t)}{N(Na~I)(t_{0})}$ and inserting it into the above equation,
we can derive a differential equation as:

\begin{equation}
\frac{dFrac(Na~I)(t)}{dt}+[R_{ion}(t)+R_{rec}(t)]\cdot Frac(Na I)(t)-R_{rec}(t)=0,
\end{equation}
Which has the following solution:
\begin{equation}
\begin{split}
Frac(Na I)(t)= \exp\{\int^{t}_{t_{exp}}-[R_{ion}(t')+R_{rec}(t')]dt'\}\times \\ \{\int^{t}_{t_{exp}}e^{\int^{t'}_{t_{exp}}[R_{ion}(t'')+R_{rec}(t'')]dt''}\cdot R_{rec}(t')dt'+1\}
\end{split}
\end{equation}

According to the photoionization theory, the ionization rate and recombination rate can be expressed as:
\begin{equation}
\begin{split}
R_{ion}(t)=\sigma\cdot I(t); \\
R_{rec}(t)=\alpha(T_{e}(t))\cdot n_{e}(t)
\end{split}
\end{equation}

Here, $\sigma$ is the photoionization cross-section of Na~I; and I(t) is the photon count rate per unit area. The photoionization cross-section depends on the energy of the incoming photons, and only the photons with energy higher than 5.139 eV can remove the electrons from the ground state for Na~I (Verner 1996a). In other words, the photons responsible for Na ionizations should have a wavelength shorter than 2412 {\AA}. Here we adopted the spectral template of SNe Ia (Nugent et~al. 2002, Hsiao et~al. 2007) to generate the UV photons needed to ionize Na~I. The UV flux has been scaled to the peak luminosity of SNe Ia with Cepheid distances (Riess et~al. 2016) and is then converted into photon count rate. Meanwhile, the second equation of Eq.(4) shows that the recombination rate of Na~II is related to the recombination coefficient $\alpha$(T$_{e}$(t))(which can change with the temperature $T$) and electron density within the CS shell n$_{e}$(t). We simply assume a constant temperature and electron density in our model by adopting them as the mean values of these two parameters. Then the recombination rate can be expressed as R$_{rec}$=$\alpha$(T$_{e}$)$\cdot$n$_{e}$. According to the studies of some literatures (i.e., Kamp et~al. 2001, Douvion et~al. 2001, Johansson et~al. 2013), we adopt a typical temperature of 100 K for the CS dust. And the relationship between $\alpha$ and T$_{e}$ is determined by (Verner et~al. 1996b).

Some model curves from our calculations, adjusted for the observed data of some SNe~Ia in our sample, are shown in Fig. 2.
One can see that the overall evolution of Na~{\sc I}D absorption seen in some HV SNe Ia is similar to that predicted by a simple CS shell model (see the solid curves in Fig.2). By comparing with the model curves, typical values of R$_{S}$$\sim$0.1 pc (3$\times$10$^{17}$ cm) and n$_{e}$$\sim$5$\times10$$^{7}$cm$^{-3}$ can be found for the subsample showing evolving Na~{\sc I}D absorptions. Of the sample listed in Table 2, the reddening towards SN 2001N may be dominated by the CS dust as Na seems to be nearly completely ionized around the peak luminosity; and the dust shell may be located at a distance with R$_{S}$$<$0.1 pc. The longer recombination time scale ($\gtrsim$30 days) suggests that the density of the CS material is quite low, e.g., n$_{e}$ ($\lesssim$10$^{7}$cm$^{-3}$). Note that estimates of the above parameters may suffer large uncertainties from the difficulties in distinguishing the CS component from the component of Na~{\sc I}D absorption in these low-resolution spectra. Moreover, it is also difficult to determine accurately the total amount of neutral Na due to line saturation. Nevertheless, the above analysis indicates that the excess reddening of HV SNe Ia is related to the CS dust.

\subsubsection{Dust Scattering Model}
Dust scattering process can be dealt with an elastic scattering (including Rayleigh scattering and Mie scattering). Light scattering in circumstellar environment of SNe Ia was first studied by Wang (2005) who found that inclusion of the scattered light tends to reduce the total extinction and hence the ratio of extinction to color excess, i.e. R$_{V}$ = A$_{V}$/E(B $-$ V). This provides an alternative explanation for the unusually low R$_{V}$ observed in some SNe Ia. Later on, Goobar (2008) conducted a similar study by considering the effects of both scattering and absorption (due to a CS shell) on the SN light. Multiple scattering process will predominately attenuate photons with shorter wavelengths, thus steepening the effective extinction law. Moreover, such a light scattering can be observed as a light echo phenomenon at late time when the SN light becomes faint enough. As blue photons usually scatter more for dust grains with smaller size, the resulting echo would thus appear blue. However, after multiple scattering, the blue photons could be finally absorbed or directed off the line of sight for some specific optical depth, while the red photons suffer less scattering and absorption. And this contrast will lead to the production of a red echo.

The extra light seen in the $B$-band light curves of the HV SNe Ia shown in Fig. 3 could be due to the scattering of SN light by the nearby CS dust, which can be quantitatively modeled for any given geometric configurations. In our models of dust scattering, we adopt Mie scattering and consider three structures of CS dust: a spherical shell, an asymmetric shell (AsyShell), and a disk configuration, as shown in Figure 5. There are more complicated configurations such as torus or blobs, and these discussions will be presented in a forthcoming paper (Hu et al. 2019 in prep.).

For the spherical shell and disk structures, the number density of dust grains along the radial direction ($N(R)$) is assumed to decrease inversely with radius squared, i.e., $A/R^2$, where $R$ is the distance to the SN and $A$ relates to the optical depth. Thus three parameters, $R_{inner}$ (inner radius of the shell), $R_{outer}$ (outer radius of the shell), and $\tau$ (optical depth) can be used to define the geometric property of the shell structure. For the asymmetric shell, however, we consider additional dependence of the dust density on the angle from the symmetry axis $\theta$, with $N(R,\theta) = A/R^2 \times (1+(\sin^n\theta - 1)s)$, where $n$ and $s$ describe degrees of asymmetry relative to the spherical shell and their best-fit values are 2.0 and 0.6, respectively. This modified relation can be regarded as an extended form of number density given in Chevalier et~al. (1986). For the disk structure, an opening angle parameter $\theta_{disk}$ is needed to address the thickness of the disk. Moreover, an observing angle $\theta_{obs}$ needs to be assumed for both the disk and AsyShell structures.

In our calculations of the dust scattering, all the values of albedo (=$\sigma_{s}$/($\sigma_{s}+\sigma_{a}$)), absorption cross-section ($\sigma_{abs}$), scattering cross-section ($\sigma_{sca}$), and phase function related to scattering process are taken from Draine et~al. (2003). The size distribution of dust grains is adopted as,
\begin{equation}
f(r) = r^{-a_0}exp\{-b_0(\log{r\over r_0})^{2.0}\}
\label{eq00}
\end{equation}
where $a_{0}$ and $b_{0}$ is adopted as $4.0$ and $7.5$, respectively, to be consistent with the results derived in Nozawa et~al.(2015).

The modeled light/color curves, obtained by taking the mean light curves of NV SNe Ia as input of light scattering, are overplotted in Figure 3. To match the observed $B$- and $V$-band light curves of our HV sample, the CS dust is required to have an inner radius of $R_{inner}$$\sim$$1\times10^{17}$ cm and an outer radius of $R_{outer}$$\sim$$2\times10^{17}cm$ for most of our HV sample, with the optical depth $\tau$ being about 0.12, 0.15, and 0.7 for the CS dust of spherical shell, asymmetric shell, and disk structures, respectively. The best-fit opening angle for the disk structure is $\theta_{disk}$ $\sim$$20^{\circ}$. For the CS dust of asymmetric shell and disk structures, the symmetry axes are tilted at an angle $\theta_{obs}\sim$30$^{\circ}$ with respect to the observer.
Note that the results listed for the above parameters represent their average values because there are many sets of parameter values that can give reasonable fit to the light curves. The corresponding mass-loss rate of the stellar wind $\dot{M}_{w}$ is estimated as $\sim$9$\times$10$^{-7}$M$_{\odot}$ yr$^{-1}$, $\sim$8$\times$ 10$^{-7}$M$_{\odot}$ yr$^{-1}$, and $\sim$6$\times$10$^{-6}$M$_{\odot}$ yr$^{-1}$ for the CS dust of shell, AsyShell, and disk structures, respectively. The calculation of $\dot{M}_{w}$ is based on the equation of N = $A/R^2 = \dot{M}_{w}/(4\pi v_w R^2 m)$, where $m$ is the average mass of each dust grain.

It is remarkable that the model light curves of different geometric configurations agree with the observed light/color scatter seen at 40-100 days after optical maximum. Note that the color measurements at these dates correspond to those used in the Lira-Phillips relation (Lira et~al. 1998) for color excess estimates of SNe Ia. This agreement is encouraging and suggestive of the validity of these models, favoring for the presence of CS dust around SNe~Ia. Moreover, the distance of the CS dust inferred from modeling of the late-time light curve is in strikingly consistent with that derived from the photoionization calculations presented in \S 3.3.1. Further distinguish between different geometric structures of the CS dust needs to reply on the polarimetric observations at late time (see more discussions below).

We notice that Bulla et al. (2018a,b) also estimated the dust locations of SNe Ia from their $B - V$ color evolution. They simulated the propagation of Monte Carlo photons through a dust region and constrained the dust distances to their SN Ia sample at
4$\times$10$^{16}$-10$^{20}$ cm, which is on average larger than our estimates. This difference is likely related to the dust properties
adopted in the analysis. In Bulla et al. analysis they chose the Mikyway-type dust and adopt a thin disk without an extended structure, while we adopt three different structures of CS dust with an average dust size of 0.5$\mu$m. A smaller dust grain is also favored for those SNe Ia with unusually low R$_{V}$ (Wang et al. 2008b, Goobar 2008). In addition, the derived color excess E$(B - V)$ from the $B - V$ color may suffer relatively large uncertainties due to larger photometric errors and that some SNe Ia may have peculiar color evolution.

\begin{figure}                                                                                                         \figurenum{5}
\includegraphics[angle=0,width=90mm]{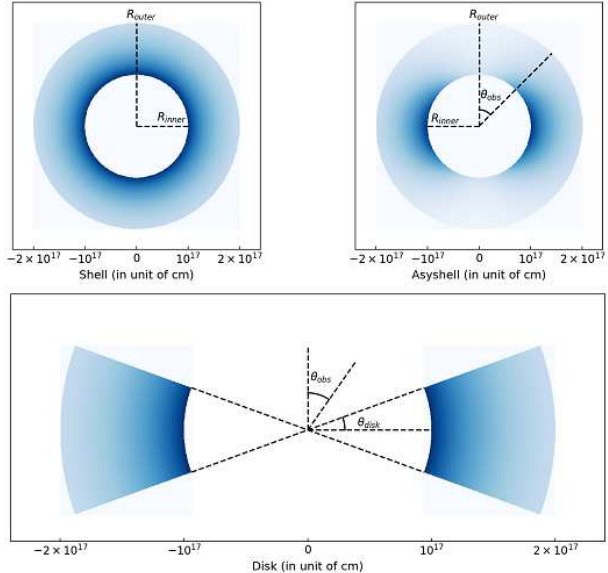}
\caption{An illustration of the vertical planes of three different structures of CS dust, spherical shell (top left), asymmetric (top right), and the disk (bottom). The vertical dashed lines show the symmetry axes of the dust distribution. $\theta_{obs}$ shows the direction to the observer. The structures are rotationally symmetric with respect to the vertical dashed lines. The boundaries of the CSM are identified by the inner radius ($R_{inner}$) and the outer radius ($R_{outer}$). The gray scale indicates the number density of the dust grains. The dust density decreases inversely with radius squared in both cases. For (a), the azimuth density dependence follows $\sin^2\theta$, with $\theta$ being the angle to the symmetry axis.}
\label{fig-5}
\end{figure}

\section{Discussions and Conclusions}
The formation of nearby CSM around an SN progenitor is a direct consequence of its progenitor evolution. The CSM can be ejected continuously similar to stellar winds, or episodically similar to nova shell ejections. Typical blueshift in the Na~{\sc I} lines of SNe~Ia is $\sim$100 km s$^{-1}$ according to a statistical study using high- and intermediate-resolution spectra of a large sample of SNe Ia (Sternberg et al. 2011; Maguire et~al. 2013). The velocity and distance of the CS dust inferred here for HV SNe~Ia suggest that they may have symbiotic nova progenitors, similar to RS Ophiuchi (Patat et al. 2011). Theoretical expectations for the fraction of SNe~Ia from the symbiotic progenitor channel are $\sim$1 to 30\% (Lu et al. 2009), which is consistent with the birthrate of HV subclass, with a fraction of about 20\% of all SNe Ia (Li et al. 2011b, Pan et al. 2015). For such a system, the fast-expanding nova shell (with a velocity of a few thousands of km s$^{-1}$) ejected from recurrent nova outbursts will be slowed down by the slow-moving stellar wind blown from the companion star, i.e., a red giant. Such an interaction process can create a large evacuated region around the SN progenitor and form a CS shell with a velocity of $\sim$100 km s$^{-1}$ at a distance ranging from $10^{15} cm$ to $10^{18} cm$, consistent with the estimates for the HV sample. The extension and distance of the CS dust from the progenitor are likely related to the accretion rate, period of recurrent nova, and the time lag between the explosion and mass loss before explosion.

\begin{figure}
\figurenum{6}
\includegraphics[angle=0,width=80mm]{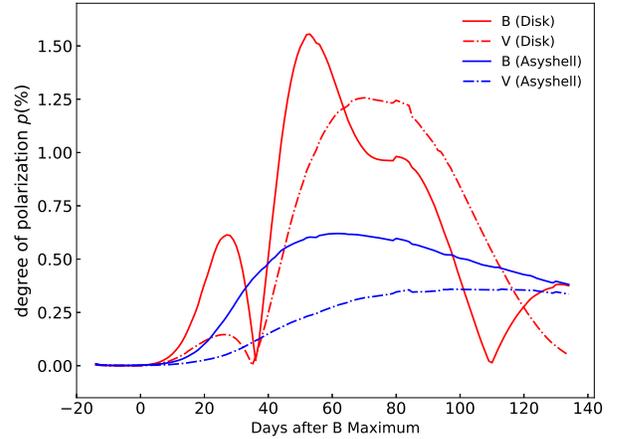}\\
\caption{The predicted time-evolution of the polarization in the $B$- (solid line) and $V$-band (dot-dashed line) at a viewing angle of 30 degrees for the asymmetric shell. The blue curves show the polarization inferred from the asymmetric shell, while the red ones represent case for the disk structure.}
\label{fig-6}
\end{figure}

The distribution of the CS dust derived for the HV SNe Ia is typically around 10$^{17}$ cm according to the above analysis, which is too compact to be spatially resolvable at extragalactic distances. The progenitors of SNe~Ia involve binaries with one or two degenerate stars, the dust ejected from such systems may be naturally asymmetric. Polarimetry study of such compact light echoes is able to map the geometric structures of the CSM and set observational constraints on the progenitor systems. We remark that the asymmetric dust distribution naturally leads to polarized radiation at late times when the scattered light is the strongest, corresponding to about 60-90 days after optical maximum, as shown in Figure 6. The absolute level of the degree of polarization is sensitive to the assumed geometric structure. Although only the SNe with highly asymmetric CS dust shells close to being edge-on are expected to produce strong polarization, the detection of such polarized signal and its evolution can be a prime evidence for the presence of the CS matter around the SNe (Wang et al. 1996). Unfortunately, existing polarimetry data are mostly taken around optical maximum (Wang \& Wheeler 2008). Indeed, recent deep HST imaging polarimetry of SN 2014J taken on day 277 after maximum reveals an abnormal polarization signal that can be identified with a light echo from a dust lump of at least 2$\times$10$^{-6}$M$_{\odot}$, located at a distance of 5$\times$10$^{17}$ cm (Yang et al. 2018). More polarimetry at late epochs coupled with photometric monitoring holds the key to unlock the mystery concerning the CSM around SN~Ia progenitors.

Our studies demonstrate that the HV SNe~Ia are associated with the CS dust, which is in accordance with the detection of systematically blueshifted velocity structure (or outflow) in the Na~I lines of SNe Ia having high Si~II velocity (Sternberg et al. 2011). There are several additional evidences favoring that the HV subclass may originate from single degenerate progenitor system. For example, a significant ultraviolet excess is possibly detected in SN 2004dt and SN 2009gi (Wang et al. 2012, Foley et al. 2013), which is expected from the collision between ejecta and a larger companion such as a massive MS star or a red giant star (Kasen 2010). Moreover, there is a tendency that unburned carbon (i.e., C~II~6580\AA) is detected in the earlier spectra of NV SNe Ia (Parrent et al. 2011, Blondin et al. 2012, Silverman et al. 2012) but not in the HV SNe Ia. Recent studies also indicate that these two subclasses of SNe Ia also show significant differences in the velocity distribution of outer-layer oxygen and silicon (Zhao et al. 2016). These results are in line with the idea that SNe~Ia with different ejecta velocities have different progenitor systems and/or explosion mechanisms. Combining with the results from the evolving narrow interstellar Na {\sc I} absorptions and the late-time light curves presented in this report, we suggest that SNe~Ia with fast expanding ejecta likely arise from progenitor systems with a red giant companion.

Nevertheless, some secular merger models may produce blueshifted absorbing materials, i.e., through interaction of tidal disrupted
ejecta with interstellar medium (Raski \& Kasen 2013) or wind blown from the accretion disk (Dragulin \& Hoflich 2016). However, the
former one can only produce an absorbing shell with a blueshifted velocity of about 100 km s$^{-1}$ , after the ejection of tidal tail for about 10$^{4}$ years when the absorbing materials moved to a distance of about 10$^{18}$ cm. While the fast wind from the accretion-disk
(with a velocity of $\sim$3000 km s$^{-1}$) tend to produce a low-density void region several light years across, surrounded by a
dense shell at a distance $\gtrsim$ 10$^{18}$ cm (Dragulin \& Hoeflich 2016). Thus, neither of these two merger models could produce the
CSM consistent with our result. Moreover, the core-degenerate (CD) scenario, an explosion of a WD and the core of an AGB star, has also
been proposed as an alternative way to explain the presence of CSM around some SNe Ia (i.e., Tsebrenko \& Soker 2015, Soker 2015), but it is not obvious whether the observed properties of a CD SN Ia would resemble that of a typical SN Ia.

\section{Acknowledgments}
We thank the anonymous referee for his/her suggestive comments to help improve the manuscript. This work is supported by the National Natural Science Foundation of China (NSFC grants 11325313, 11633002, and 11761141001), and the National Program on Key Research and Development Project (grant no. 2016YFA0400803). The work of L.W. is supported by NSF grant AST-1817099. This research has made use of the CfA Supernova Archive, which is funded in part by the National Science Foundation through grant AST 0907903. This research has also made use of the supernova archives from the Carnegie Supernova Project and the Berkeley Supernova Program, which are also funded in part by the US National Science Foundation.

\clearpage
\LongTables
\begin{deluxetable}{llccccccr}
\hspace{-1.0cm}
\tablecolumns{9}\tablewidth{0pc} \tabletypesize{\scriptsize}
\tablecaption{Summary of Classification, Spectroscopic and Photometric Properties of the SN~Ia Sample}
\tablehead{\colhead{SN name} & \colhead{SN type} & \colhead{v$^{0}_{Si~II}$} & \colhead{$\Delta$m$_{15}$(B)}
&\colhead{$B_{max} - V_{max}$} & \colhead{EW} & \colhead{$\Delta$m$_{60}$(B)} & \colhead{$\Delta$m$_{60}$(V)} & \colhead{Ref.\tablenotemark{$\ast$}} \\
\colhead{}&\colhead{} &\colhead{($\times10^{4}$ km s$^{-1}$)} & \colhead{(mag)} & \colhead{(mag)} & \colhead{(\AA)} & \colhead{(mag)}
& \colhead{(mag)} & \colhead{}}  \\
\startdata
\multicolumn{9}{c}{}\\
\noalign{\smallskip}
\hline
&       &      &          &  Branch normal         &         &          &         &   \\
\hline
SN 1984A  & HV   & 1.47(04) & 1.22(10) &  0.16(09)     & ...     &3.10(10) & 2.59(05) & 1   \\
SN 1989B  & NV   & 1.05(03) & 1.35(05) &  0.32(09)     &3.17(19) &3.41(06) & ...      & 2   \\
SN 1994M  & HV   & 1.24(02) & 1.47(06) &  0.10(04)     &0.22(21) & ...     & ...      & 3,4,5 \\
SN 1994ae & NV   & 1.11(03) & 0.89(05) &$-$0.05(04)    &0.29(06) &3.46(03) & 2.60(03) & 3,4,6 \\
SN 1995D  & NV   & 1.01(02) & 1.01(03) &  0.01(03)     &0.33(09) &3.32(03) & 2.47(03) & 3,4,5 \\
SN 1995E  & NV   & 1.08(02) & 1.20(07) &  0.69(05)     &1.97(39) & ...     & 2.49(05) & 3,4,5 \\
SN 1995al & HV   & 1.25(03) & 0.95(05) &  0.09(04)     &1.04(06) &3.25(05) & 2.48(06) & 4,5 \\
SN 1996C  & NV   & 1.08(03) & 0.95(03) &  0.09(04)     & ...     &3.23(05) & 2.58(04) & 4,5 \\
SN 1996X  & NV   & 1.13(03) & 1.30(05) &$-$0.01(03)    &0.00(12) &3.55(05) & 2.72(04) & 3,4,5 \\
SN 1996Z  & NV   & 1.17(03) & 1.18(18) &  0.26(07)     &1.51(17) & ...     & ...      & 4,5 \\
SN 1996ai & NV   & 1.06(02) & 1.00(08) &  1.47(07)     &$>$2.8   & ...     & ...      & 3,4,5 \\
SN 1996bo & HV   & 1.23(05) & 1.31(06) &  0.36(04)     &1.63(16) & ...     & ...      & 4,5 \\
SN 1997E  & NV   & 1.18(03) & 1.42(05) &  0.05(03)     &0.10(20) &3.60(03) & 2.83(05) & 4,7 \\
SN 1997Y  & NV   & 1.11(03) & 1.39(05) &  0.01(03)     &1.23(21) &  ...    & ...      & 4,7\\
SN 1997bp & HV   & 1.57(05) & 1.11(03) &  0.21(03)     &1.85(09) &3.13(05) & 2.48(03) & 4,7 \\
SN 1997bq & HV   & 1.45(05) & 1.13(10) &  0.10(04)     &0.51(07) & ...     & ...      & 4,7 \\
SN 1997do & HV   & 1.25(05) & 1.08(03) &  0.13(04)     &0.17(26) &3.30(07) & 2.46(05) & 7,8 \\
SN 1998V  & NV   & 1.09(03) & 0.98(04) &  0.03(03)     &0.26(10) & ...     & ...      & 3,7,8 \\
SN 1998aq & NV   & 1.07(02) & 1.02(03) &$-$0.08(03)    &0.16(08) &3.51(02) & 2.62(02) & 6,8 \\
SN 1998bu & NV   & 1.08(03) & 1.06(04) &  0.33(03)     &0.45(10) &3.34(03) & 2.60(04) & 8,9 \\
SN 1998dh & NV   & 1.18(02) & 1.22(04) &  0.07(03)     &0.36(12) &3.33(03) & 2.64(03) & 8,10 \\
SN 1998dk & HV   & 1.26(03) & 1.15(03) &  0.17(06)     &1.15(17) & ...     & ...      & 3,4 \\
SN 1998dm & NV   & 1.02(03) & 0.96(05) &  0.27(05)     &0.62(13) &3.34(03) & 2.55(03) & 8,10 \\
SN 1998ec & HV   & 1.26(03) & 1.18(04) &  0.22(07)     &1.07(23) & ...     & ...      & 8,10 \\
SN 1998ef & HV   & 1.24(05) & 1.36(06) &$-$0.04(04)    &0.31(12) &3.40(04) & 2.62(03) & 4,10 \\
SN 1998eg & NV   & 1.05(03) & 1.21(06) &  0.02(03)     &1.49(18) & ...     & ...      & 6,8,10 \\
SN 1999cl & HV   & 1.22(03) & 1.26(05) &  1.08(04)     &3.45(08)  & ...     & ...      & 4,8,10 \\
SN 1999cp & NV   & 1.09(03) & 1.01(03) &  0.02(03)     &0(08)     & ...     & ...      & 3,4,10 \\
SN 1999dk & HV   & 1.26(06) & 1.15(05) &  0.08(03)     &0.10(10)  &3.30(04) & 2.65(04) & 3,10 \\
SN 1999ej & NV   & 1.07(03) & 1.57(08) &  0.04(03)     &0(11)     &3.53(06) & 2.80(06) & 8,10 \\
SN 2000cn & NV   & 1.15(05) & 1.68(06) &  0.12(03)     &0(15)     &3.46(10) & 2.84(06) & 8,10 \\
SN 2000cp & NV   & 1.15(03) & 1.24(04) &  0.41(04)     &2.64(19)  & ...     & ...      & 3,4,10 \\
SN 2000cw & NV   & 1.06(03) & 1.33(05) &  0.06(03)     &0(14)     &...      & ...      & 3,4,7 \\
SN 2000dk & NV   & 1.09(03) & 1.66(03) &  0(03)        &0.10(14)  &3.64(07) & 3.00(03) & 3,8,10 \\
SN 2000dm & NV   & 1.12(03) & 1.51(04) &  0.02(03)     &0(06)     &...      & ...      & 3,4,10 \\
SN 2000dn & NV   & 1.03(03) & 1.13(03) &$-$0.03(03)    &0(20)     &...      & ...      & 3,4,10 \\
SN 2000dr & NV   & 1.05(05) & 1.73(10) &  0.11(04)     & ...      &3.73(13) & 3.07(10) & 3,10 \\
SN 2000fa & NV   & 1.18(03) & 0.98(09) &  0.11(03)     &0.97(19)  &3.29(04) & 2.58(03) & 3,8,7,10 \\
SN 2001E  & HV   & 1.40(05) & 1.08(05) &  0.43(04)     &1.44(24)  & ...     & ....     & 4,10 \\
SN 2001ay & HV   & 1.44(02) & 0.68(05) &  0.06(06)     &0.71(15)  & ...     & ...      & 4,10 \\
SN 2001bf & NV   & 1.10(03) & 0.83(03) &  0.03(03)     &0(12)     &3.00(07) & 2.23(04) & 3,4,10 \\
SN 2001bg & HV   & 1.23(03) & 1.14(03) &  0.15(03)     &1.02(09)  & ...     & ...      & 4,10 \\
SN 2001br & HV   & 1.35(03) & 1.32(06) &  0.10(03)     &1.10(21)  & ...     & ...      & 4,10 \\
SN 2001ck & NV   & 1.15(04) & 1.09(04) &$-$0.01(03)    &1.24(20)  & ...     & ...      & 4,10 \\
SN 2001cp & NV   & 1.07(03) & 0.90(04) &  0.04(03)     &0.63(18)  & ...     & ...      & 4,10 \\
SN 2001da & NV   & 1.17(03) & 1.30(04) &  0.17(03)     &0.71(12)  &3.26(06) & 2.54(03) & 4,10 \\
SN 2001en & HV   & 1.30(03) & 1.23(03) &  0.04(02)     &0.98(14)  &3.35(05) & 2.69(04) & 4,10 \\
SN 2001ep & NV   & 1.05(03) & 1.40(05) &  0.08(02)     &0.73(20)  &3.55(05) & 2.69(05) & 4,10 \\
SN 2001fe & NV   & 1.09(02) & 0.96(10) &$-$0.03(03)    &0.71(16)  &3.39(05) & 2.57(03) & 3,11 \\
SN 2002aw & NV   & 1.06(02) & 1.15(03) &  0.08(03)     &1.06(21)  &3.50(07) & 2.51(04) & 3,4,10 \\
SN 2002bf & HV   & 1.61(04) & 1.08(10) &  0.17(05)     &1.18(24)  &3.14(05) & 2.57(04) & 4,10 \\
SN 2002bo & HV   & 1.32(02) & 1.15(03) &  0.40(08)     &2.44(09)  &3.18(03) & 2.56(02) & 4,10 \\
SN 2002cd & HV   & 1.50(03) & 0.94(06) &  0.63(04)     &1.71(10)  &2.96(10) & 2.51(04) & 3,4,10,11\\
SN 2002cr & NV   & 0.98(02) & 1.19(05) &  0(02)        &0.51(12)  &3.45(04) & 2.60(03) & 4,10 \\
SN 2002cs & HV   & 1.36(06) & 1.07(05) &  0.12(03)     & ...      &3.18(04) & 2.10(03) & 4,10 \\
SN 2002cu & HV   & 1.23(05) & 1.48(04) &  0.07(03)     &0(20)     &3.48(06) & 2.84(04) & 3,4,10 \\
SN 2002de & NV   & 1.13(02) & 1.05(03) &  0.16(03)     &1.60(32)  &3.40(08) & 2.70(06) & 4,10 \\
SN 2002dj & HV   & 1.34(02) & 1.13(03) &  0.07(03)     &0.80(13)  &3.29(10) & 2.57(08) & 4,10 \\
SN 2002dp & NV   & 1.07(02) & 1.28(03) &  0.07(02)     &0.70(14)  &3.45(04) & 2.62(03) & 4,10 \\
SN 2002eb & NV   & 1.03(02) & 0.93(03) &$-$0.03(03)    &0(20)     &3.34(04) & 2.44(06) & 3,10 \\
SN 2002ef & HV   & 1.16(04) & 1.18(05) &  0.32(04)     &1.99(32)  &3.36(09) & 2.64(04) & 3,4,10 \\
SN 2002er & NV   & 1.17(03) & 1.32(03) &  0.12(03)     &1.75(09)  &3.42(05) & 2.65(04) & 10,12 \\
SN 2002fk & NV   & 0.98(03) & 1.02(03) &$-$0.10(03)    &0.07(08)  &3.43(03) & 2.61(03) & 4,10 \\
SN 2002ha & NV   & 1.09(03) & 1.37(03) &$-$0.02(03)    &0.83(16)  & ...     & 2.82(29) & 3,4,10 \\
SN 2002he & NV   & 1.16(05) & 1.41(05) &  0.02(03)     &0(10)     &3.53(07) & 2.78(10) & 3,4,10 \\
SN 2002hu & NV   & 1.04(03) & 1.02(03) &$-$0.07(03)    & ...      & ...     & 2.45(21) & 3,4,11 \\
SN 2002jg & NV   & 1.09(03) & 1.49(07) &  0.67(06)     &2.70(50)  & ...     & ...      & 4,10 \\
SN 2003W  & HV   & 1.51(04) & 1.15(03) &  0.16(04)     &0.99(20)  &3.05(09) & 2.54(06) & 4,10,11 \\
SN 2003cg & NV   & 1.09(03) & 1.17(04) &  1.14(04)     &5.43(23)  &3.39(03) & 2.62(03) & 4,10,11 \\
SN 2003du & NV   & 1.04(03) & 1.00(02) &$-$0.06(03)    &0.10(07)  &3.38(03) & 2.60(03) & 4,10,11,13 \\
SN 2003gn & HV   & 1.23(03) & 1.30(08) &  0.10(03)     & ...      &3.59(10) & 2.90(10) & 3,4,10 \\
SN 2003hv & NV   & 1.13(03) & 1.45(07) &$-$0.08(04)    &0(10)     &3.59(04) & 2.92(03) & 3,4,10 \\
SN 2003kf & NV   & 1.11(03) & 0.93(04) &  0.02(03)     &0.50(10)  &3.37(10) & 2.57(03) & 3,4,10,11 \\
SN 2004L  & NV   & 1.02(02) & 1.42(10) &  0.23(05)     &1.15(23)  &  ...    & ...      & 3,4,11 \\
SN 2004S  & NV   & 0.97(03) & 1.08(07) &$-$0.04(05)    &0(20)     &3.45(05) & 2.61(04) & 10,14 \\
SN 2004as & HV  & 1.21(03)  & 1.20(05) &  0.09(03)     &0.26(25)  &3.41(05) & 2.59(03) & 3,4,10,11 \\
SN 2004at & NV  & 1.09(02)  & 1.07(03) &$-$0.06(03)    &0.19(14)  &3.46(06) & 2.64(03) & 4,10 \\
SN 2004bg & NV  & 1.04(03)  & 1.12(07) &$-$0.08(03)    &0(20)     &3.54(03) & 2.76(06) & 4,10 \\
SN 2004bk & HV  & 1.22(05)  & 0.98(08) &  0.06(04)     &0.47(12)  &3.34(06) & 2.63(06) & 3,4,10 \\
SN 2004dt & HV  & 1.39(03)  & 1.10(05) &$-$0.01(03)    &0.46(21)  &3.15(05) & 2.51(05) & 10,15 \\
SN 2004ef & HV  & 1.20(05)  & 1.42(03) &  0.09(03)     &0.90(27)  &3.36(04) & 2.75(04) & 4,16 \\
SN 2004eo & NV  & 1.07(03)  & 1.40(03) &  0.03(03)     &0(20)     &3.48(04) & 2.69(04) & 16,17 \\
SN 2004ey & NV  & 1.11(03)  & 1.03(03) &$-$0.10(03)    &0(30)     &3.44(04) & 2.47(05) & 3,10,16,18 \\
SN 2004fz & NV  & 1.02(03)  & 1.34(04) &  0.01(03)     &0.30(20)  & ...     & ...      & 4,10 \\
SN 2004gs & NV  & 1.08(03)  & 1.50(05) &  0.15(02)     &1.47(15)  &3.56(04) & 2.95(03) & 4,16,18 \\
SN 2004gu & NV  & 1.10(03)  & 0.82(03) &  0.12(03)     &1.63(16)  & ...     & ...      & 16,18 \\
SN 2005A  & HV  & 1.42(03)  & 1.19(05) &  1.00(03)     &3.50(20)  & ...     & ...      & 16,18 \\
SN 2005W  & NV  & 1.08(02)  & 1.19(03) &  0.14(03)     &0.76(15)  & ...     &  ...     & 16,18 \\
SN 2005ag & NV  & 1.13(03)  & 1.04(05) &$-$0.02(03)    & ...      &3.19(04) & 2.64(03) & 16,18 \\
SN 2005al & NV  & 1.09(03)  & 1.23(05) &$-$0.07(03)    &0(10)     &3.53(03) & 2.82(05) & 16,18 \\
SN 2005am & NV  & 1.18(02)  & 1.45(06) &  0.04(03)     &0.21(10)  &3.50(05) & 2.87(05) & 4,10,16 \\
SN 2005bc & NV  & 1.08(03)  & 1.45(06) &  0.43(05)     &1.90(25)  &3.53(06) & 2.86(07) & 3,4,10 \\
SN 2005bg & NV  & 1.07(02)  & 0.99(05) &  0(03)        &1.07(11)  & ...     & ...      & 16,18 \\
SN 2005bo & NV  & 1.09(02)  & 1.30(05) &  0.26(04)     &1.38(14)  &3.58(10) & 2.70(06) & 4,10,16,18 \\
SN 2005cf & NV  & 1.01(03)  & 1.07(03) &  0.01(03)     &0.30(06)  &3.42(03) & 2.54(04) & 19 \\
SN 2005de & NV  & 1.03(03)  & 1.19(05) &  0.08(03)     &0.51(10)  &3.47(07) & 2.62(04) & 3,10 \\
SN 2005el & NV  & 1.05(02)  & 1.30(03) &$-$0.09(03)    &0.20(18)  &3.46(05) & 2.80(06) & 3,4,10,16 \\
SN 2005hc & NV  & 1.13(03)  & 0.96(05) &$-$0.01(03)    & ...      &3.35(04) & 2.64(05) & 4,16,18 \\
SN 2005iq & NV  & 1.10(05)  & 1.30(03) &$-$0.09(03)    &0(24)     &3.62(06) & 2.87(06) & 3,4,16 \\
SN 2005kc & NV  & 1.04(02)  & 1.16(10) &  0.25(04)     &1.78(13)  & ...     & ...      & 3,16,18 \\
SN 2005ki & NV  & 1.10(02)  & 1.43(06) &$-$0.09(03)    &0.30(09)  &3.54(04) & 2.87(03) & 3,16,18 \\
SN 2005lu & HV  & 1.37(10)  & 0.97(03) &  0.19(03)     &2.02(32)  &3.21(05) & 2.56(04) & 16,18\\
SN 2005ms & NV  & 1.13(03)  & 1.02(07) &$-$0.03(03)    &0(20)     &3.38(05) & 2.67(11) & 3,11 \\
SN 2005na & NV  & 1.04(02)  & 1.08(07) &$-$0.08(03)    &0(23)     &3.48(03) & 2.62(02) & 3,4,16 \\
SN 2006D  & NV  & 1.08(02)  & 1.33(06) &  0.03(03)     &0(05)     &3.50(03) & 2.91(02) & 3,4,16,18 \\
SN 2006N  & NV  & 1.14(02)  & 1.52(08) &  0.01(03)     &0.82(21)  &3.94(10) & 2.97(05) & 3,4,11 \\
SN 2006S  & NV  & 1.10(02)  & 0.91(04) &  0.04(03)     &0.46(09)  & ...     & 2.50(10) & 3,4,11 \\
SN 2006X  & HV  & 1.61(02)  & 1.26(05) &  1.23(04)     &2.05(08)  &2.97(03) & 2.48(02) & 11,16,20 \\
SN 2006ac & HV  & 1.32(06)  & 1.23(08) &  0.04(03)     &0(13)     &3.40(08) & 2.72(05) & 4,11 \\
SN 2006ax & NV  & 1.01(02)  & 1.00(04) &$-$0.08(02)    &0(12)     &3.43(05) & 2.55(04) & 3,4,11,16 \\
SN 2006cp & HV  & 1.27(06)  & 1.09(05) &  0.06(03)     &0.26(14)  & ...     & ...      & 3,4,10,11 \\
SN 2006ef & HV  & 1.23(03)  & 1.37(06) &$-$0.01(03)    &0.50(10)  &3.52(05) & 2.86(05) & 3,10 \\
SN 2006ej & HV  & 1.22(05)  & 1.38(07) &  0(05)        &0(23)     &3.41(05) & 2.82(05) & 3,10,21 \\
SN 2006en & NV  & 1.06(03)  & 1.12(05) &  0.10(03)     &1.22(24)  & ...     & ...      & 3,10 \\
SN 2006et & NV  & 0.97(03)  & 0.93(03) &  0.16(02)     &5.19(23)  &3.35(04) & 2.55(03) & 3,4,21 \\
SN 2006gr & HV  & 1.34(07)  & 0.91(05) &  0.11(03)     &1.48(23)  & ...     & 2.61(08) & 4,10,11 \\
SN 2006hb & NV  & 1.00(03)  & 1.55(09) &  0.04(03)     &0.10(21)  &3.63(04) & 2.93(04) & 4,18,21 \\
SN 2006is & HV  & 1.38(05)  & 0.63(04) &$-$0.05(03)    &0.23(20)     &3.02(04) & 2.45(03) & 18,21 \\
SN 2006le & NV  & 1.18(06)  & 0.89(05) &$-$0.03(03)    &0.10(08)  &3.26(07) & 2.56(04) & 4,10,11 \\
SN 2006os & NV  & 1.16(03)  & 1.23(07) &  0.32(03)     &1.85(37)  & ...     & ...      & 3,18,21 \\
SN 2006sr & NV  & 1.18(03)  & 1.38(06) &  0.01(03)     &0.58(12)  & ...     & ...      & 3,4,11 \\
SN 2007A  & NV  & 1.08(02)  & 0.95(03) &  0.14(03)     &0.90(14)  & ...     & ...      & 3,4,11,21 \\
SN 2007F  & NV  & 1.11(03)  & 0.95(03) &$-$0.06(03)    &0.65(17)  &3.45(07) & 2.66(05) & 3,4,11 \\
SN 2007af & NV  & 1.08(02)  & 1.16(03) &  0.01(03)     &0.31(08)  &3.43(03) & 2.61(03) & 3,4,11,21 \\
SN 2007as & HV  & 1.33(03)  & 1.24(03) &  0.09(03)     & ...      &3.37(03) & 2.73(05) & 18,21 \\
SN 2007bc & NV  & 1.02(02)  & 1.29(05) &  0.01(03)     &0.62(19)  &3.70(10) & 2.90(05) & 3,4,21 \\
SN 2007bd & HV  & 1.24(05)  & 1.32(05) &$-$0.04(03)    &0.45(09)  & ...     & ...      & 3,4,21 \\
SN 2007bm & NV  & 1.05(05)  & 1.27(04) &  0.41(03)     &2.17(07)  & ...     & ...      & 3,4,11,21 \\
SN 2007ca & NV  & 1.05(03)  & 0.97(04) &  0.26(03)     &1.94(14)  &3.40(06) & 2.67(06) & 3,4,11,21 \\
SN 2007ci & NV  & 1.17(02)  & 1.77(09) &  0.04(03)     &0.10(15)  & ...     & ...      & 3,4,10,11 \\
SN 2007co & NV  & 1.18(02)  & 1.13(05) &  0.12(03)     &0(16)     & ...     & ...      & 3,4,10 \\
SN 2007cq & NV  & 1.05(03)  & 1.23(03) &$-$0.03(03)    &0(14)     &3.45(11) & 2.68(08) & 3,4,10,11 \\
SN 2007fb & NV  & 1.14(03)  & 1.39(05) &$-$0.09(03)    &0.62(22)  &3.65(05) & 2.90(05) & 3,4,23 \\
SN 2007gi & HV  & 1.56(03)  & 1.37(05) &  0.14(03)     &0.23(10)  &3.25(03) & 2.70(03) & 22 \\
SN 2007hj & HV  & 1.20(02)  & 1.64(05) &  0.12(03)     &0.36(09)  &3.62(14) & 2.93(07) & 4,10,22 \\
SN 2007jg & HV  & 1.26(04)  & 1.22(03) &  0.07(03)     & ...      &3.46(04) & 2.87(04) & 4,22,23 \\
SN 2007kk & HV  & 1.25(05)  & 0.87(04) &$-$0.01(03)    &0.53(19)  &3.08(34) & 2.40(15) & 4,23 \\
SN 2007le & HV  & 1.29(06)  & 1.02(05) &  0.29(03)     &1.52(06)  &3.16(03) & 2.56(02) & 3,4,10,23 \\
SN 2007nq & NV  & 1.19(02)  & 1.54(04) &  0(02)        &0(20)     &3.63(05) & 2.97(03) & 4,18,21,23\\
SN 2007qe & HV  & 1.35(05)  & 0.98(03) &  0.05(03)     &0.11(07)  &3.30(06) & 2.70(07) & 3,4,10,11 \\
SN 2007sr & HV  & 1.31(05)  & 1.05(07) &  0.08(03)     &0.31(12)  &3.20(03) & 2.57(02) & 4,10,16,18 \\
SN 2007ss & NV  & 1.18(03)  & 1.35(05) &  0.28(04)     &2.75(19)  &3.55(10) & 2.69(08) & 4,23 \\
SN 2007sw & HV  & 1.34(03)  & 1.04(05) &  0.12(04)     &1.00(13)  & ...     &  ...     & 4,23 \\
SN 2007ux & NV  & 1.07(03)  & 1.53(08) &  0.11(03)     &0(15)     &3.63(05) & 2.95(04) & 3,4,18 \\
SN 2008C  & NV  & 1.06(03)  & 1.19(03) &  0.15(03)     &1.22(08)  & ...     & 2.69(04) & 4,16,18,23 \\
SN 2008Z  & NV  & 1.12(03)  & 0.88(03) &  0.10(03)     &0.93(21)  &3.30(05) & 2.60(04) & 3,4,23 \\
SN 2008ar & NV  & 1.02(02)  & 1.09(03) &$-$0.02(03)    &0(19)     &3.38(03) & 2.62(03) & 3,4,10,21 \\
SN 2008bc & NV  & 1.16(03)  & 0.92(03) &$-$0.04(03)    &0.35(07)  &3.31(04) & 2.63(02) & 3,4,21 \\
SN 2008bf & NV  & 1.13(02)  & 0.95(04) &$-$0.08(03)    &0.08(15)  &3.40(03) & 2.64(03) & 3,4,10,11,21 \\
SN 2008bq & NV  & 1.05(02)  & 1.01(03) &  0.09(02)     &0.55(11)  & ...     & ...      & 18,21 \\
SN 2008dr & HV  & 1.44(03)  & 1.52(06) &  0.08(03)     &0.38(13)  & ...     & ....     & 3,10 \\
SN 2008ec & NV  & 1.10(02)  & 1.33(04) &  0.15(03)     &0.61(12)  &3.62(10) & 2.71(06) & 3,10 \\
SN 2008fp & NV  & 1.12(03)  & 0.89(03) &  0.40(03)     &2.42(20)  &3.30(03) & 2.54(02) & 18,21 \\
SN 2008fr & NV  & 1.04(04)  & 0.95(10) &$-$0.07(03)    &0(20)     &3.41(03) & 2.46(03) & 21,23,24 \\
SN 2008gl & HV  & 1.20(03)  & 1.35(04) &  0.03(03)     &0(13)     &3.46(04) & 2.80(03) & 18,21 \\
SN 2008gp & NV  & 1.06(02)  & 1.05(04) &$-$0.09(03)    &0(14)     & ...     & ...      & 18,21 \\
SN 2008hv & NV  & 1.09(03)  & 1.21(05) &$-$0.07(03)    &0(19)     &3.50(02) & 2.84(02) & 18,21,23 \\
SN 2008ia & NV  & 1.13(03)  & 1.31(05) &$-$0.02(03)    &0.93(31)  & ...     & ...      & 18,21 \\
SN 2009Y  & HV  & 1.47(04)  & 0.97(03) &  0.15(03)     &0.15(05)  &3.01(03) & 2.51(02) & 18,21,23 \\
SN 2009aa & NV  & 1.05(03)  & 1.21(03) &$-$0.07(03)    &0(19)     &3.60(03) & 2.71(03) & 18,21 \\
SN 2009ab & NV  & 1.08(03)  & 1.22(04) &$-$0.02(03)    &0.51(21)  & ...     & ...      & 18,21 \\
SN 2009ad & NV  & 1.03(02)  & 1.00(03) &$-$0.05(03)    &0.90(15)  & ...     & ...      & 18,21,23 \\
SN 2009ag & NV  & 1.02(02)  & 1.06(05) &   0.17(03)    &0.40(08)  & ...     & ...      & 18,21 \\
SN 2009ig & HV  & 1.30(03)  & 0.90(05) &   0.07(03)    &0.52(14)  &3.05(03) & 2.48(03) & 3,23 \\
SN 2011fe & NV  & 1.04(02)  & 1.18(03) &$-$0.03(03)    &0.07(05)  &3.55(03) & 2.65(03) & 25 \\
SN 2014J  & HV  & 1.21(02)  & 0.98(02) &   1.20(03)    &5.27(26)  &3.18(03) & 2.44(02) & 26,27 \\
\hline
\hline
          &     &           &          & Peculiar 91T-like Subclass &      &         &          &   \\
\hline
SN 1991T  & 91T   & 0.98(02) & 0.94(03) & 0.18(03)     &1.30(14) &3.12(03) &2.43(03)  &28,29 \\
SN 1995ac & 91T   & 0.95(03) & 0.83(05) &$-$0.01(05)   &2.04(53) &3.16(04) & 2.55(07) &4,5 \\
SN 1995bd & 91T   & 0.96(03) & 0.87(03) & 0.42(03)     &1.49(15) &3.05(04) & 2.43(05) & 4,5 \\
SN 1998ab & 91T   & 1.00(04) & 1.03(02) & 0.07(03)     &0.60(19) &3.45(05) & 2.51(03) & 4,5 \\
SN 1998es & 91T   & 1.04(02) & 0.78(03) & 0.09(03)     &1.36(12) &3.25(05) & 2.38(04) & 8,10 \\
SN 1999aa & 91T   & 1.05(02) & 0.81(02) &$-$0.01(02)   &0.05(12) &3.20(07)& 2.45(04) & 8,10 \\
SN 1999ac & 91T   & 1.00(03) & 1.30(03) & 0.09(03)     &0.05(08) &3.38(02)& 2.51(02) & 7,8,10 \\
SN 1999dq & 91T   & 1.08(03) & 0.90(03) & 0.07(03)     &1.00(07) &3.27(04) & 2.46(03) & 4,6,9 \\
SN 1999gp & 91T   & 1.10(03) & 0.82(03) & 0.04(03)     &0.27(20) & ...     & ...      & 4,5,8 \\
SN 2000cx & 91T   & 1.14(03) & 0.98(03) &-0.10(03)     &0(20)    &3.52(03) & 3.12(02) & 3,4,8 \\
SN 2001V  & 91T   & 1.13(03) & 0.92(03) & -0.01(03)    &1.26(11)  &3.21(03) & 2.45(03) & 4,5 \\
SN 2001eh & 91T   & 1.03(02) & 0.78(03) & 0.02(03)     &0(20)     &3.11(04) & ...      & 3,4 \\
SN 2003fa & 91T   & 1.03(03) & 0.88(03) &-0.03(03)     & ...      &3.62(08) & 2.51(07) & 4,10\\
SN 2004bv & 91T   & 0.71(03) & 0.98(03) &0.10(04)      &0.71(13)  &3.22(06) & 2.42(03) & 3,10 \\
SN 2004br & 91T   & 1.10(04) & 0.87(03) &-0.04(03)     &0(19)     & ...     & ...      & 3,10 \\
SN 2004gu & 91T   & 1.14(02) & 0.80(06) &0.13(03)      &1.83(15)  & ...     & ...      & 16,17,18\\
SN 2005M  & 91T   & 0.81(03) & 0.85(05) &-0.01(03)     &0.25(13)  &3.31(03)  &2.51(03) & 3,10 \\
SN 2005eq & 91T   & 1.00(02) & 0.78(03) &0.04(03)      &0.75(17)  &3.25(05) & 2.52(02) & 4,11 \\
SN 2005eu & 91T   & 1.04(05) & 0.87(03) &-0.08(03)     &0.60(28)  & ...     & ...      & 3,4,10,11\\
SN 2005ls & 91T   & 1.10(03) & 0.95(10) &0.30(03)      &1.81(36)  & ...     & 2.52(03) & 5,6 \\
SN 2006cm & 91T   & 1.10(02) & 1.10(10) &0.94(04)      &2.12(39)  & ...     & ...      & 4,11 \\
SN 2006mp & 91T   & 0.81(03) & 0.92(03) &-0.01(03)     &0(20)     &3.51(13) & ...      & 4,11 \\
SN 2007S  & 91T   & 0.97(03) & 0.89(03) &0.38(03)      &1.70(18)  &3.22(05) & 2.38(03) & 4,11 \\
SN 2007ai & 91T   & 0.97(03) & 0.90(03) &0.20(03)      &1.46(26)  &3.28(09) & 2.55(05) & 16,18 \\
SN 2007cq & 91T   & 1.04(02) & 1.14(09) &0.04(03)      & ...      &3.25(04) & ...      & 4,11 \\
\hline
\hline
&         &       &          & Peculiar 91bg-like Subclass &      &         &          &   \\
\hline
SN 1991bg & 91bg  & 1.02(02) & 1.93(05) & 0.75(05)     & ...      &3.19(08) & ...      & 30  \\
SN 1998bp & 91bg  & 1.07(02) & 1.83(05) & 0.30(10)     &0(14)     &3.52(10) & 3.02(05) & 4,7 \\
SN 1998de & 91bg  & 1.11(03) & 1.96(05) & 0.71(03)     &0(30)     &3.24(10) & 3.06(10) & 4,10 \\
SN 1999by & 91bg  & 1.02(03) & 1.91(05) & 0.50(03)     &0(20)     &3.51(10) & 3.20(10) & 4,8 \\
SN 1999da & 91bg  & 1.13(02) & 1.94(05) & 0.68(05)     &0(16)     & ...     & ...      & 10 \\
SN 2002cf & 91bg  & 1.07(02) & 1.86(10) & 0.47(03)     & ...      & ...     & 3.14(08) & 3,10 \\
SN 2002dl & 91bg  & 1.15(05) & 1.85(03) & 0.14(03)     &0.54(17)  &3.53(06) & 2.94(05) & 3,4,10 \\
SN 2002fb & 91bg  & 1.07(03) & 1.88(05) & 0.41(04)     &0.31(11)  &3.30(12) & 3.17(10) & 3,4,10 \\
SN 2003gs & 91bg  & 1.14(03) & 1.93(07) & 0.63(07)     & ...      &3.25(06) & 3.15(07) & 3,10 \\
SN 2005bl & 91bg  & 0.98(02) & 1.93(10) & 0.78(03)     &2.60(30)  & ...     & 3.16(17) & 31 \\
SN 2005ke & 91bg  & 0.95(03) & 1.81(04) & 0.61(03)     &0(10)     &3.11(03)	& 3.07(04) & 4,16 \\
SN 2005mz & 91bg  & 1.07(05) & 1.96(07) & 0.33(07)     &...       &3.50(15)  &3.26(07) & 4,11 \\
SN 2006bz & 91bg  & 1.09(03) & 2.09(16) & 0.70(08)     &0.43(33)  & ...     & ...      & 4,11 \\
SN 2006em & 91bg  & 1.03(03) & 1.98(16) & 0.89(10)     &0(32)     & ...     & ...      & 4,11 \\ 
SN 2008bt & 91bg  & 0.98(03) & 1.90(10) & 0.47(03)     & ...      & 3.33(04) &3.00(05) & 16,18,21

\enddata
\tablenotetext{}{Note: The uncertainties shown in the brackets are 1$\sigma$, in units of 0.01 mag for $\Delta$m$_{15}$(B), B$_{max}$ $-$ V$_{max}$, $\Delta$m$_{60}$(B), and $\Delta$m$_{60}$(V), and in units of 0.01 \AA\ for EW of Na~I D absorption.}
\tablenotetext{$\ast$}{1= Barbon et al. 1989; 2 = Wells et al. 1994; 3 = Silverman et al. 2012; 4 = Blondin et al. 2012; 5 = Riess et al. 1999; 6 = Riess et al. 2005; 7 = Jha et al. 2006; 8 = Matheson et al. 2008; 9 = Jha et al. 1999; 10 = Ganeshelingam et al. 2010; 11 = Hicken et al. 2009; 12 = Kotak et al. 2005; 13 = Stanishev et al. 2007; 14 = Krisciunas et al. 2007; 15 = Altavilla et al. 2007; 16 = Contreras et al. 2010, Krisciunas et al. 2017; 17 = Pastorello et al. 2007; 18 = Folatelli et al. 2013; 19 = Wang et al. 2009; 20 = Wang et al. 2008a; 21 = Stritzinger et al. 2011, Krisciunas et al. 2017; 22 = Zhang et al. 2010; 23 = Hicken et al. 2012; 24 = Yuan et al. 2008; 25 = Zhang et al. 2016; 26 = Brown et~al. 2015; 27 = Zhang et al. 2018; 28= Filippenko 1992; 29 = Lira et al. 1998; 30 = Filippenko et~al. 1992; 31 = Taubenberger et~al. 2008.}
\end{deluxetable}

\clearpage
\begin{deluxetable}{lcclll}
\hspace{-1.0cm}
\tablecolumns{5}\tablewidth{0pc} \tabletypesize{\scriptsize}
\tablecaption{Candidates of type Ia supernovae with variable Na I D absorption features}
\tablehead{\colhead{SN name} & \colhead{$\Delta$EW(\AA)\tablenotemark{a}} & \colhead{$\Delta$EW/$\sigma$\tablenotemark{b}} &
\colhead{Number\tablenotemark{c}} &\colhead{Phase of the Spectra (days)\tablenotemark{d}}  & \colhead{Type}} \\
\startdata
SN 1997bp & 0.66 & 3.5 & 13 &-1.6,2.3,+20.3, +33.2,+51.2 & HV  \\
SN 1997bq & 0.96 & 4.7 & 8 &-10.6, -4.5, +12.4, +20.4 & HV   \\
SN 1999cl & 1.07 & 5.0 & 11& -6.7,-2.6, +9.3, +39.3 & HV \\
SN 1999dq & 0.73 & 3.8 & 21 & -9.5, -3.5, 30.3, +59.3,+90.3 & 91T \\
SN 2001br & 1.27 & 4.4 & 4 &-1.1,+1.9, +26.9,+53.8 & HV  \\
SN 2001N  & 1.64 & 6.9 & 6 &+3.6,+10.6,+12,7, +33.5 & HV   \\
SN 2002bf & 1.38 & 4.7 & 4 &+3.8,+6.8,+8.8,+12.8 & HV  \\
SN 2002bo & 0.85 & 3.4 & 31 &-12.0,-5.9,-1.1,+11.0,+22.0& HV  \\
SN 2002cd & 2.19 & 12.5 & 9 &-8.7,-4.7,+0.3,+15.2, & HV  \\
SN 2002dj & 1.89 & 10.0 & 6 &-10.0,-5.9,+10.1,+17.1& HV  \\
SN 2003ep & 2.23 & 10.6 & 3 & a few weeks later & \nodata  \\
SN 2005hf & 1.10 & 3.9 & 6 &+3.0,+7.0,+13.1 & Pec   \\
SN 2006N & 1.02 & 4.1 & 7 &+2.2,+5.2,+8.2,+28.2 & NV \\
SN 2006X & 0.93 & 8.2 & 21 &-6.6,+3.0,+8.0,+15.0,+34.0 & HV  \\
SN 2006dd &2.53 & 3.7 & 5  &-12.0, +86.4, +137.0, +194.4 & \nodata \\
SN 2007le & 0.67 & 4.0 & 25 &-6.3,+5.6,+19.6,+40.5,+67.6& HV \\
SN 2008C & 1.27 & 6.5 & 5 &+9.5,+13.5,+30.5,+43.5 & NV  \\
\enddata
\tablenotetext{a}{The maximum variation of the equivalent width (EW) measured from the Na I absorption in the spectra of SNe Ia.}
\tablenotetext{b}{The confidence that the SNe Ia show variable Na~ID absorption, obtained by dividing the maximum variation of EW over the unertainty in the measurement.}
\tablenotetext{c}{The number of the spectra used in the analysis.}
\tablenotetext{d}{Relative to the B-band maximum, and only the representative phases are shown for the limited space.}
\end{deluxetable}

\end{document}